\documentclass[12pt]{article}

\usepackage{amsmath}
\usepackage{graphicx,psfrag,epsf}
\usepackage{enumerate}
\usepackage{natbib}

\usepackage[utf8]{inputenc}
\usepackage{amsthm}
\usepackage{mathtools}
\usepackage{multirow}
\usepackage{caption}
\usepackage{subcaption}
\usepackage{color}
\usepackage{float} 

\newtheorem*{mypropo*}{Proposition}

\newtheorem*{mycoro*}{Corollary}

\newcommand{\blind}{0}

\begin{document}

\def\spacingset#1{\renewcommand{\baselinestretch}%
{#1}\small\normalsize} \spacingset{1}



\if0\blind
{
  \title{\bf A Quantile Estimate Based on Local Curve Fitting}
  \author{Dr. Mar\'ia I. Salazar-Alvarez \\
    School of Engineering and Sciences, Tecnologico de Monterrey\\
    \\  
    Dr. V\'ictor G. Tercero-G\'omez\\
    School of Engineering and Sciences, Tecnologico de Monterrey\\
	\\  
	Dr. Alvaro E. Cordero-Franco \\
    Facultad de Ciencias F\'isico Matem\'aticas, Universidad Autonoma de Nuevo Leon\\
\\  
	Dr. William J. Conover
    \\
    Department of Mathematics and Statistics, Texas Tech University\\
    \\
    Dr. Mario G. Beruvides
    \\
    Department of Industrial Engineering, Texas Tech University\\
    \\
  }
  \maketitle
} \fi

\if1\blind
{
  \bigskip
  \bigskip
  \bigskip
  \begin{center}
    {\LARGE\bf A Quantile Estimate Based on Local Curve Fitting}
\end{center}
  \medskip
} \fi

\bigskip

\begin{abstract}
Quantile estimation is a problem presented in fields such as quality control, hydrology, and economics. There are different techniques to estimate such quantiles. Nevertheless, these techniques use an overall fit of the sample when the quantiles of interest are usually located in the tails of the distribution. Regression Approach for Quantile Estimation (RAQE) is a method based on regression techniques and the properties of the empirical distribution to address this problem. The method was first presented for the problem of capability analysis. In this paper, a generalization of the method is presented, extended to the multiple sample scenario, and data from real examples is used to illustrate the proposed approaches. In addition, theoretical framework is presented to support the extension for multiple homogeneous samples and the use of the uncertainty of the estimated probabilities as a weighting factor in the analysis. 
\end{abstract}

\noindent%

\noindent%
\vfill

\newpage
\spacingset{2} 

\section{Introduction}
\label{sec:intro}
Quantile estimation is a problem presented in different fields such as quality control, risk analysis, finance, and  hydrology, among other fields. Sometimes it is necessary to know the value of a certain quantile of a given distribution in order to use certain statistical tools. For these purposes, various approaches for quantile estimation have been proposed in the literature based on curve-fitting techniques, resampling techniques, or transformations.    

Curve fitting techniques require the first four moments of a sample in order to estimate the quantiles of interest by using some known curve or probability distribution. Some of the most used curves are the Pearson family of curves, the Gumbel distribution, and the lognormal distribution, just to mention a few. \cite{daniel1999fitting} in their book present the strengths and limitations of the method of least squares when using curve fitting techniques. The main problem with this approach is that usually the quantiles of interest are located at the extreme of the distribution, and this technique requires information from the center of the distribution. The center values get a good fit  while the adjustment at the tails gets compromised.

Transformation techniques use different models to transform the original distribution to a normal distribution. Some of the most used transformations are those proposed by \cite{box1964analysis} and \cite{johnson1949systems}. Other authors such as \cite{bartlett1947use}, \cite{velleman1981applications}, and \cite{tan2004using} present different transformations and conclusions of when to use each of them. Nevertheless, a major disadvantage with these techniques is that the quantiles estimates depend on the knowledge of the right transformation. Several transformations must be tried first before finding a useful one. 

Resampling techniques use different subsets of data taken randomly with replacement from a sample of interest. By using different subsets, the parameter of interest, in this case a quantile of interest, can be estimated. This resampling usually uses bootstrap and jackknife methods. Additional information about these methods can be found in \cite{davison1997bootstrap}, \cite{efron1982jackknife}, and \cite{efron1983leisurely}, among others. However, resampling techniques are known to require additional computational complexity, making the approach awkward and unwieldy for industry practitioners.

All these previous methods have shown acceptable performance for different situations, however, all of the known methods require an overall fit on the available data, giving the same weights to all observations when usually the quantiles of interest are located in the tails of the distribution.The only approaches found in the literature that deals with different weights when fitting a distribution are the weighted maximum likelihood approaches. Nevertheless, they are parametric in nature and need to assume a specific probability distribution. For instance, \cite{field1994robust} uses this approach to reduce the "pull" given by outliers when performing an estimation. \cite{salazar2016regressing} presented a regression approach to improve the estimates when dealing with extreme values. Regression Approach for Quantile Estimation, RAQE,  was introduced to solve the problem of process capability estimation based on the percentile method. In their paper, the authors show a series of Monte Carlo simulations to demonstrate the performance of the method, and a case study was presented in the automotive related industry.  The problem was limited to the use of one sample and the estimation of process capability indexes. 

This paper aims to generalize the methodology proposed by \cite{salazar2016regressing}. Since RAQE is based on regression techniques it  can be easily confused with research areas like quantile regression and extremal quantile regression; nevertheless,  RAQE does not use variable conditioning. It only assumes a set of independent and identically distributed observations. Moreover, the problem of the uncertainty of a probability is incorporated into the regression techniques to improve the curve estimation. In addition, an extension of the methodology is presented for its use in the case of multiple homogeneous samples. 

In this paper the generalization of RAQE is presented in Section 2, for one or multiple homogeneous samples. Examples to illustrate the applicability of the method are presented as well.  Section 3 provides a theoretical discussion that supports the proposed approach. Finally, in Section 4, general conclusions about the proposed method and suggested future work are provided.

\section{Methods}
\label{sec:meth}
This section presents a methodology to estimate quantiles by using local curve fitting for single and multiple homogeneous samples. Some examples to illustrate the use of the proposed methodology are presented as well.

\subsection{RAQE}
\label{subsec:RAQE}
Let $X=\{X_{(i)}:i=1, \ldots, n\}$ be a random vector of ordered statistics from a sample of independent and identically distributed observations, where $x=\{x_i:i,\ldots,n \}$ corresponds to the observed values. This implies that  $x_1 < \ldots < x_n$ . Then, to estimate the desired extreme quantiles use the following procedure:

\begin{enumerate}

\item Augment the empirical distribution function (e.d.f)
\begin{equation}
S_n(x)=\frac{1}{n}\sum_{i=1}^{n} {I_{(-\infty, x]}(X_i)}, \qquad -\infty < x < \infty,
\end{equation}
with additional points to assist in fitting a continuous function to approximate the discrete function $S_n(x)$. The additional points are defined as
\begin{align*}
(a_{2i}, b_{2i})=((X_{(i)}+X_{(i+1)})/2, i/n), \qquad  i=1,2,\ldots, n-1, 
\end{align*}
and
\begin{align*}
(a_{2i-1}, b_{2i-1})=(X_{(i)}, (i-1/2)/n), \qquad  i=1,2,\ldots, n. 
\end{align*}
If the quantile of interest is located on the lower tail; then, choose a family of curves $g(x \mid \theta)$ to fit to the ordered augmented set, $\{a_i, b_i\}_{i=1}^{2m-1}$ for a selected value of $m < n/2$.
Use a search method to find the set of parameters $\theta$ that minimizes the weighted least square criterion 
\begin{equation}
\sum_{i=1}^{2m-1}{w_i(b_{(i)}-g(x \mid \theta))^2}
\end{equation}
where $w_i=\frac{1}{\hat{\sigma}_i^2}$, correspond to the weights given to each of the observations. For the case of the empirical distribution the weights are estimated as $w_i=n/(b_i(1-b_i))$.

If the quantile of interest is located on the upper tail, then choose a family of curves  $h(x \mid \varphi)$ to fit the ordered augmented set $\{a_i, b_i\}_{2n-2l+1}^{2n-1}$ for a selected value of $l<n/2$. Use a search method to find the set of parameters $\varphi$ that minimizes
\begin{equation}
\sum_{i=2n-1l+1}^{2n-1}{w_i(b_{(i)}-h(x \mid \varphi))^2}
\end{equation} 
\item Estimate the quantiles of interest using the inverse image of the augmented empirical distribution, $\hat{X}_{p}= g^{-1}(p)$ for the lower tail and/or $\hat{X}_{p}= h^{-1}(p)$ for the upper tail, where p represents the $p^{th}$ quantile of the data. 
\end{enumerate}

\subsubsection{The Case of Multiple Homogeneous Samples}
\begin{figure}[h]
    \begin{center}
    \includegraphics[width=.75\textwidth]{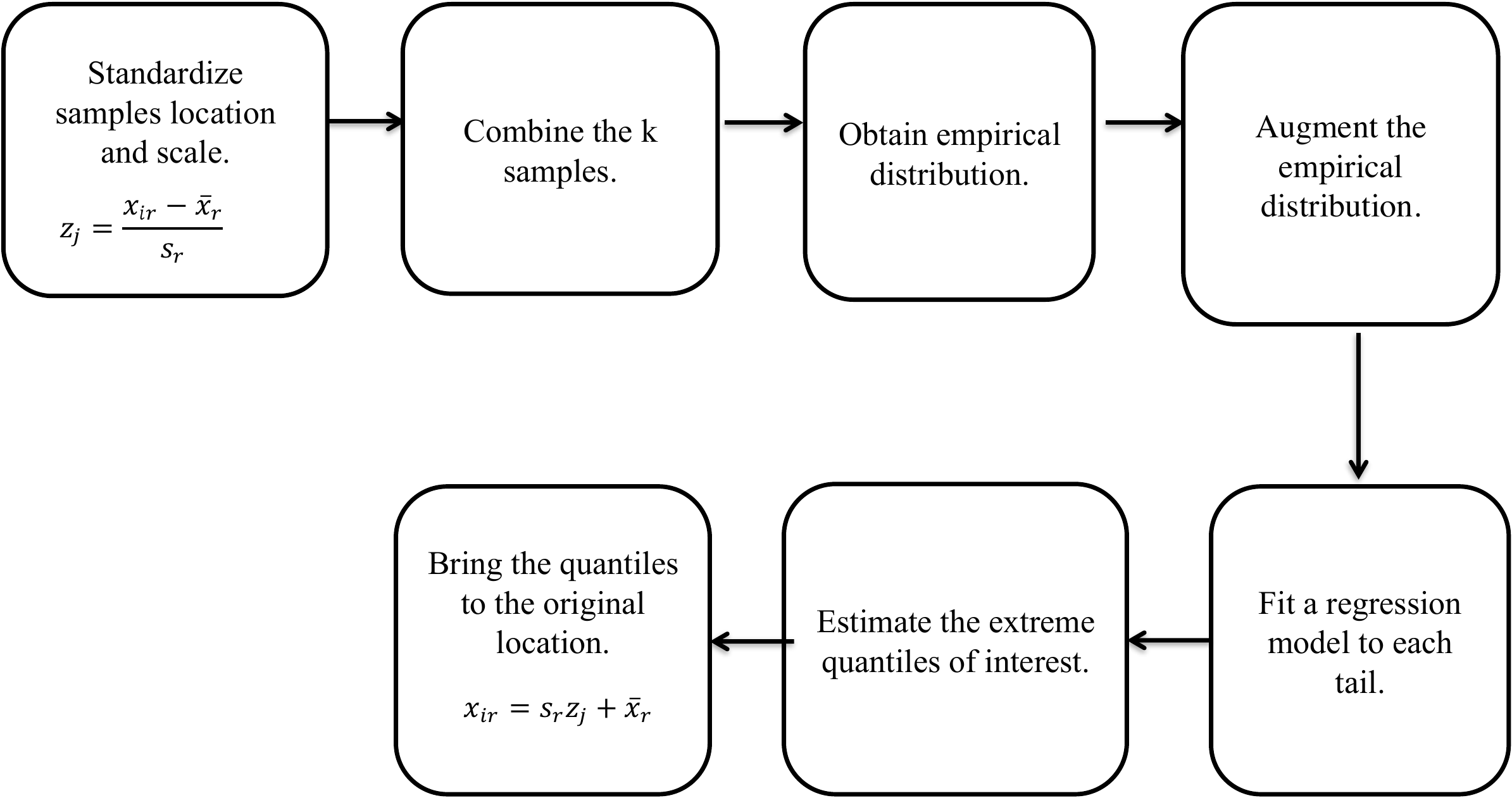}
        \caption{Flow Diagram for RAQE for Multiple Homogeneous Samples}
    \label{Graph 1}
   	\end{center}
\end{figure}

RAQE could also be used in the case of multiple homogeneous samples. Homogeneous samples are defined as multiple samples that have the same distribution but differ on their location and/or scale parameters. These types of samples are present in industry when dealing with lines that produce similar products, for instance, a bottling company that uses the same line to fill $375$ ml and $500$ ml bottles. For these scenarios, the problem of multiple samples is reduced to the problem of one sample by combining the k-homogeneous samples into one big sample by standardizing with the corresponding sample mean and sample standard deviation, centering and re-scaling whenever there is a significance difference found in location or scale. After the quantiles of interest are obtained from the combined sample, the estimation needs to be return to the original form $x_{ir}= s_{r}z_j + \bar{x}_{r}$, where $\bar{x}_{r}$ is the sample mean, $s_{r}$ is the sample standard deviation, $r$ stands for the individual samples $\left(r=1,2,\ldots, k \right)$, and $i$ represents the observation within samples  $\left(i=1,2,\ldots, n_k \right)$ and $j$ stands for the standardize observation $\left(j=1,2,\ldots,\sum_{r=1}^{k} n_r \right)$. For clarity, the method is presented in Figure ~\ref{Graph 1}.


\subsection{Real Data Examples}
In this section, two examples are presented using real data to demonstrate the applicability of the methodology in real scenarios. 

\subsubsection{Estimation of Control Limits}
\label{subsec:Example 1}
To illustrate the method an example given by \cite{chou1998control} is shown in this section. Measurements of particle count in wafers were recorded from a piece of equipment used in the senmiconductor industry.  A random  sample of size $116$ is taken from the collected data, see Table \ref{Table 1}. The authors seek to estimate the control limits of a control chart; nevertheless, the distribution of the sample is found to be skewed. Therefore, the proposed solution is to try different transformations and use the one with the higher p-value. This will be compared with the RAQE method. 

\begin{table}[h!]
\centering
\caption{Particle Count Found per Wafer}
\label{Table 1}
\begin{tabular}{|c|c|c|c|c|c|c|c|cccc}
\hline
27 & 16 & 16 & 34 & 14 & 13 & 10 & 43 & \multicolumn{1}{c|}{23} & \multicolumn{1}{c|}{13} & \multicolumn{1}{c|}{8}  & \multicolumn{1}{c|}{22} \\ \hline
14 & 8  & 15 & 20 & 79 & 9  & 13 & 6  & \multicolumn{1}{c|}{18} & \multicolumn{1}{c|}{13} & \multicolumn{1}{c|}{34} & \multicolumn{1}{c|}{16} \\ \hline
15 & 11 & 4  & 9  & 12 & 9  & 38 & 7  & \multicolumn{1}{c|}{15} & \multicolumn{1}{c|}{7}  & \multicolumn{1}{c|}{4}  & \multicolumn{1}{c|}{31} \\ \hline
9  & 7  & 35 & 7  & 8  & 15 & 13 & 5  & \multicolumn{1}{c|}{4}  & \multicolumn{1}{c|}{4}  & \multicolumn{1}{c|}{13} & \multicolumn{1}{c|}{7}  \\ \hline
39 & 61 & 27 & 11 & 10 & 18 & 14 & 3  & \multicolumn{1}{c|}{15} & \multicolumn{1}{c|}{14} & \multicolumn{1}{c|}{8}  & \multicolumn{1}{c|}{12} \\ \hline
9  & 13 & 35 & 11 & 23 & 11 & 9  & 11 & \multicolumn{1}{c|}{42} & \multicolumn{1}{c|}{12} & \multicolumn{1}{c|}{4}  & \multicolumn{1}{c|}{4}  \\ \hline
15 & 9  & 8  & 10 & 11 & 25 & 10 & 19 & \multicolumn{1}{c|}{8}  & \multicolumn{1}{c|}{19} & \multicolumn{1}{c|}{11} & \multicolumn{1}{c|}{13} \\ \hline
12 & 37 & 44 & 12 & 9  & 11 & 74 & 12 & \multicolumn{1}{c|}{27} & \multicolumn{1}{c|}{43} & \multicolumn{1}{c|}{4}  & \multicolumn{1}{c|}{6}  \\ \hline
5  & 15 & 3  & 24 & 22 & 10 & 23 & 16 & \multicolumn{1}{c|}{5}  & \multicolumn{1}{c|}{40} & \multicolumn{1}{c|}{27} & \multicolumn{1}{c|}{16} \\ \hline
5  & 12 & 23 & 5  & 30 & 19 & 8  & 9  & \multicolumn{1}{l}{}    & \multicolumn{1}{l}{}    & \multicolumn{1}{l}{}    & \multicolumn{1}{l}{}    \\ \cline{1-8}
\end{tabular}
\end{table}

\begin{figure}[h]
	\begin{center}
    \includegraphics[width=.85\textwidth]{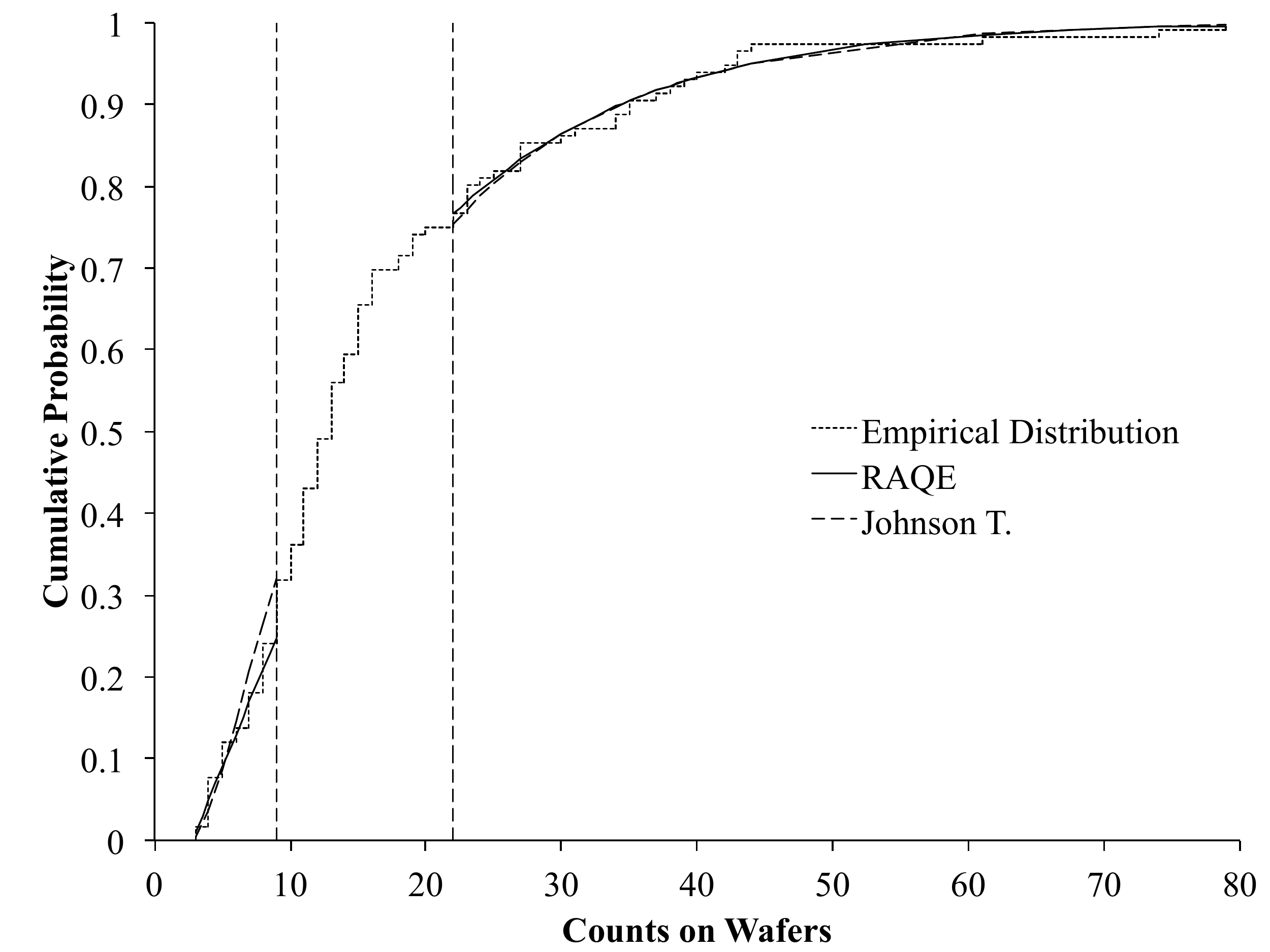}
    \caption{Empirical Distribution of the Data}
    \label{Graph 2}
    \end{center}
\end{figure}

For this data set, the Johnson transformation, the logarithmic transformation, and the square root transformation were evaluated. The selected transformation method was the Johnson transformation. The sample was then transformed and the upper and lower control limits were estimated with the moving range. The estimated values for the upper and lower control limits using transformations were  $82.1991$ and $2.8146$. For the nonparametric approach, the quantiles of interest to estimate the lower and upper control limit are $0.00135$ and $0.99865$, which is equivalent of having a range of $6\sigma$ under normality. The size of $m$ and $l$ was set to work with the  $25\%$ of the observed lower and the upper tails, respectively, and avoid the information located at the center of the distribution where a fit is not required. The distribution used in the sample for the lower tail was a quadratic function; while on the upper tail the Gumbel cumulative distribution function was used. The estimated values obtained with RAQE were $92.3982$ and $2.8022$. It can be seen that the estimated values using both methods are relatively close. This can also be seen in Figure \ref{Graph 2}. As can be seen, the fit at the tails is not as good as the one obtained with RAQE. The mean square error (MSE) for the fit on the tails using Johnson transformation was of $0.107$ and $0.0119$ for the lower and upper tail, respectively; meanwhile, using RAQE the MSE was of $0.012$ and $0.006$.    

\subsubsection{Estimation of Return Periods}
\label{subsec:Example 2}
Flood frequency estimation is an approach used in hydrology to analyze extreme weather events. Its main concern is to estimate the probability of the occurrence of a rare event in order to estimate the risk to the population and avoid death and monetary damages related to harsh weather. One of the main steps needed is to determine the frequency distribution that better fits the data under study in order to estimate the quantile related to the return period. If different data sets are found to be  homogeneous in terms of shape, but they might have different location or scale parameters, samples might be combined adequately to increase the accuracy of the curve fitting estimation. This process can be also achieved using RAQE for multiple homogeneous samples. 

The data set used in this example was obtained from climatological stations that were defined as homogenous by \cite{lara2014multivariate}.  
The information was obtained by the authors using the software Rapid Extractor of Climatological Information version $3$, but it could also be obtained from google earth by installing a software application from the Comisi\'on Nacional del Agua (CONAGUA) web site. The example uses records from the annual maximum precipitation from a region located in Sinaloa, at the northwest of Mexico. The two climatological stations are $25081$, located in the city of Culiacan, and $25078$, located in the city of Rosario. The period of time selected was from $1963$ to $2006$. The data from the two stations are presented in Table \ref{Table 2}.  

\begin{table}[h]
\centering
\caption{Annual Maximum Precipitation for Stations 25081 and 25078}
\label{Table 2}
\centering
\small
\begin{tabular}{lcccccccccc}
\cline{2-11}
\multicolumn{1}{l|}{}                        & \multicolumn{1}{c|}{\textbf{1963}} & \multicolumn{1}{c|}{\textbf{1964}} & \multicolumn{1}{c|}{\textbf{1965}} & \multicolumn{1}{c|}{\textbf{1966}} & \multicolumn{1}{c|}{\textbf{1967}} & \multicolumn{1}{c|}{\textbf{1968}} & \multicolumn{1}{c|}{\textbf{1969}} & \multicolumn{1}{c|}{\textbf{1970}} & \multicolumn{1}{c|}{\textbf{1971}} & \multicolumn{1}{c|}{\textbf{1972}} \\ \hline
\multicolumn{1}{|l|}{\textbf{Station 25081}} & \multicolumn{1}{c|}{158}           & \multicolumn{1}{c|}{72.5}          & \multicolumn{1}{c|}{123}           & \multicolumn{1}{c|}{58.1}          & \multicolumn{1}{c|}{90}            & \multicolumn{1}{c|}{96.2}          & \multicolumn{1}{c|}{76.8}          & \multicolumn{1}{c|}{128.6}         & \multicolumn{1}{c|}{101.8}         & \multicolumn{1}{c|}{188}           \\ \hline
\multicolumn{1}{|l|}{\textbf{Station 25078}} & \multicolumn{1}{c|}{79.5}          & \multicolumn{1}{c|}{160}           & \multicolumn{1}{c|}{252.8}         & \multicolumn{1}{c|}{129.5}         & \multicolumn{1}{c|}{94.8}          & \multicolumn{1}{c|}{240}           & \multicolumn{1}{c|}{80.8}          & \multicolumn{1}{c|}{193}           & \multicolumn{1}{c|}{86}            & \multicolumn{1}{c|}{268}           \\ \hline
                                             & \multicolumn{1}{l}{}               & \multicolumn{1}{l}{}               & \multicolumn{1}{l}{}               & \multicolumn{1}{l}{}               & \multicolumn{1}{l}{}               & \multicolumn{1}{l}{}               & \multicolumn{1}{l}{}               & \multicolumn{1}{l}{}               & \multicolumn{1}{l}{}               & \multicolumn{1}{l}{}               \\ \cline{2-11} 
\multicolumn{1}{l|}{}                        & \multicolumn{1}{c|}{\textbf{1973}} & \multicolumn{1}{c|}{\textbf{1974}} & \multicolumn{1}{c|}{\textbf{1975}} & \multicolumn{1}{c|}{\textbf{1976}} & \multicolumn{1}{c|}{\textbf{1977}} & \multicolumn{1}{c|}{\textbf{1978}} & \multicolumn{1}{c|}{\textbf{1979}} & \multicolumn{1}{c|}{\textbf{1980}} & \multicolumn{1}{c|}{\textbf{1981}} & \multicolumn{1}{c|}{\textbf{1982}} \\ \hline
\multicolumn{1}{|l|}{\textbf{Station 25081}} & \multicolumn{1}{c|}{62}            & \multicolumn{1}{c|}{74.5}          & \multicolumn{1}{c|}{62.5}          & \multicolumn{1}{c|}{136}           & \multicolumn{1}{c|}{94.3}          & \multicolumn{1}{c|}{82.5}          & \multicolumn{1}{c|}{56}            & \multicolumn{1}{c|}{84.5}          & \multicolumn{1}{c|}{55}            & \multicolumn{1}{c|}{75}            \\ \hline
\multicolumn{1}{|l|}{\textbf{Station 25078}} & \multicolumn{1}{c|}{111.5}         & \multicolumn{1}{c|}{109}           & \multicolumn{1}{c|}{85}            & \multicolumn{1}{c|}{87}            & \multicolumn{1}{c|}{95}            & \multicolumn{1}{c|}{53}            & \multicolumn{1}{c|}{78}            & \multicolumn{1}{c|}{60}            & \multicolumn{1}{c|}{98}            & \multicolumn{1}{c|}{87}            \\ \hline
                                             & \multicolumn{1}{l}{}               & \multicolumn{1}{l}{}               & \multicolumn{1}{l}{}               & \multicolumn{1}{l}{}               & \multicolumn{1}{l}{}               & \multicolumn{1}{l}{}               & \multicolumn{1}{l}{}               & \multicolumn{1}{l}{}               & \multicolumn{1}{l}{}               & \multicolumn{1}{l}{}               \\ \cline{2-11} 
\multicolumn{1}{l|}{}                        & \multicolumn{1}{c|}{\textbf{1983}} & \multicolumn{1}{c|}{\textbf{1984}} & \multicolumn{1}{c|}{\textbf{1985}} & \multicolumn{1}{c|}{\textbf{1986}} & \multicolumn{1}{c|}{\textbf{1987}} & \multicolumn{1}{c|}{\textbf{1988}} & \multicolumn{1}{c|}{\textbf{1989}} & \multicolumn{1}{c|}{\textbf{1990}} & \multicolumn{1}{c|}{\textbf{1991}} & \multicolumn{1}{c|}{\textbf{1992}} \\ \hline
\multicolumn{1}{|l|}{\textbf{Station 25081}} & \multicolumn{1}{c|}{58.2}          & \multicolumn{1}{c|}{82.5}          & \multicolumn{1}{c|}{72.1}          & \multicolumn{1}{c|}{224.3}         & \multicolumn{1}{c|}{89}            & \multicolumn{1}{c|}{98}            & \multicolumn{1}{c|}{116.4}         & \multicolumn{1}{c|}{127.2}         & \multicolumn{1}{c|}{83.3}          & \multicolumn{1}{c|}{82.9}          \\ \hline
\multicolumn{1}{|l|}{\textbf{Station 25078}} & \multicolumn{1}{c|}{77.7}          & \multicolumn{1}{c|}{81}            & \multicolumn{1}{c|}{90}            & \multicolumn{1}{c|}{200}           & \multicolumn{1}{c|}{152}           & \multicolumn{1}{c|}{168}           & \multicolumn{1}{c|}{64}            & \multicolumn{1}{c|}{113}           & \multicolumn{1}{c|}{66}            & \multicolumn{1}{c|}{157}           \\ \hline
                                             & \multicolumn{1}{l}{}               & \multicolumn{1}{l}{}               & \multicolumn{1}{l}{}               & \multicolumn{1}{l}{}               & \multicolumn{1}{l}{}               & \multicolumn{1}{l}{}               & \multicolumn{1}{l}{}               & \multicolumn{1}{l}{}               & \multicolumn{1}{l}{}               & \multicolumn{1}{l}{}               \\ \cline{2-11} 
\multicolumn{1}{l|}{}                        & \multicolumn{1}{c|}{\textbf{1993}} & \multicolumn{1}{c|}{\textbf{1994}} & \multicolumn{1}{c|}{\textbf{1995}} & \multicolumn{1}{c|}{\textbf{1996}} & \multicolumn{1}{c|}{\textbf{1997}} & \multicolumn{1}{c|}{\textbf{1998}} & \multicolumn{1}{c|}{\textbf{1999}} & \multicolumn{1}{c|}{\textbf{2000}} & \multicolumn{1}{c|}{\textbf{2001}} & \multicolumn{1}{c|}{\textbf{2002}} \\ \hline
\multicolumn{1}{|l|}{\textbf{Station 25081}} & \multicolumn{1}{c|}{81.3}          & \multicolumn{1}{c|}{103}           & \multicolumn{1}{c|}{53.9}          & \multicolumn{1}{c|}{114.1}         & \multicolumn{1}{c|}{63.4}          & \multicolumn{1}{c|}{60.4}          & \multicolumn{1}{c|}{83.6}          & \multicolumn{1}{c|}{72}            & \multicolumn{1}{c|}{85.9}          & \multicolumn{1}{c|}{73.8}          \\ \hline
\multicolumn{1}{|l|}{\textbf{Station 25078}} & \multicolumn{1}{c|}{115}           & \multicolumn{1}{c|}{103.5}         & \multicolumn{1}{c|}{57.5}          & \multicolumn{1}{c|}{91}            & \multicolumn{1}{c|}{63}            & \multicolumn{1}{c|}{202}           & \multicolumn{1}{c|}{122}           & \multicolumn{1}{c|}{120}           & \multicolumn{1}{c|}{170}           & \multicolumn{1}{c|}{77}            \\ \hline
                                             & \multicolumn{1}{l}{}               & \multicolumn{1}{l}{}               & \multicolumn{1}{l}{}               & \multicolumn{1}{l}{}               & \multicolumn{1}{l}{}               & \multicolumn{1}{l}{}               & \multicolumn{1}{l}{}               & \multicolumn{1}{l}{}               & \multicolumn{1}{l}{}               & \multicolumn{1}{l}{}               \\ \cline{2-5}
\multicolumn{1}{l|}{}                        & \multicolumn{1}{c|}{\textbf{2003}} & \multicolumn{1}{c|}{\textbf{2004}} & \multicolumn{1}{c|}{\textbf{2005}} & \multicolumn{1}{c|}{\textbf{2006}} & \multicolumn{1}{l}{}               & \multicolumn{1}{l}{}               & \multicolumn{1}{l}{}               & \multicolumn{1}{l}{}               & \multicolumn{1}{l}{}               & \multicolumn{1}{l}{}               \\ \cline{1-5}
\multicolumn{1}{|l|}{\textbf{Station 25081}} & \multicolumn{1}{c|}{70.9}          & \multicolumn{1}{c|}{89.4}          & \multicolumn{1}{c|}{94.2}          & \multicolumn{1}{c|}{116.9}         & \multicolumn{1}{l}{}               & \multicolumn{1}{l}{}               & \multicolumn{1}{l}{}               & \multicolumn{1}{l}{}               & \multicolumn{1}{l}{}               & \multicolumn{1}{l}{}               \\ \cline{1-5}
\multicolumn{1}{|l|}{\textbf{Station 25078}} & \multicolumn{1}{c|}{81}            & \multicolumn{1}{c|}{85}            & \multicolumn{1}{c|}{119}           & \multicolumn{1}{c|}{88}            & \multicolumn{1}{l}{}               & \multicolumn{1}{l}{}               & \multicolumn{1}{l}{}               & \multicolumn{1}{l}{}               & \multicolumn{1}{l}{}               & \multicolumn{1}{l}{}               \\ \cline{1-5}
\end{tabular}
\end{table}

\begin{figure}[h]
	\begin{center}
    \includegraphics[width=.65\textwidth]{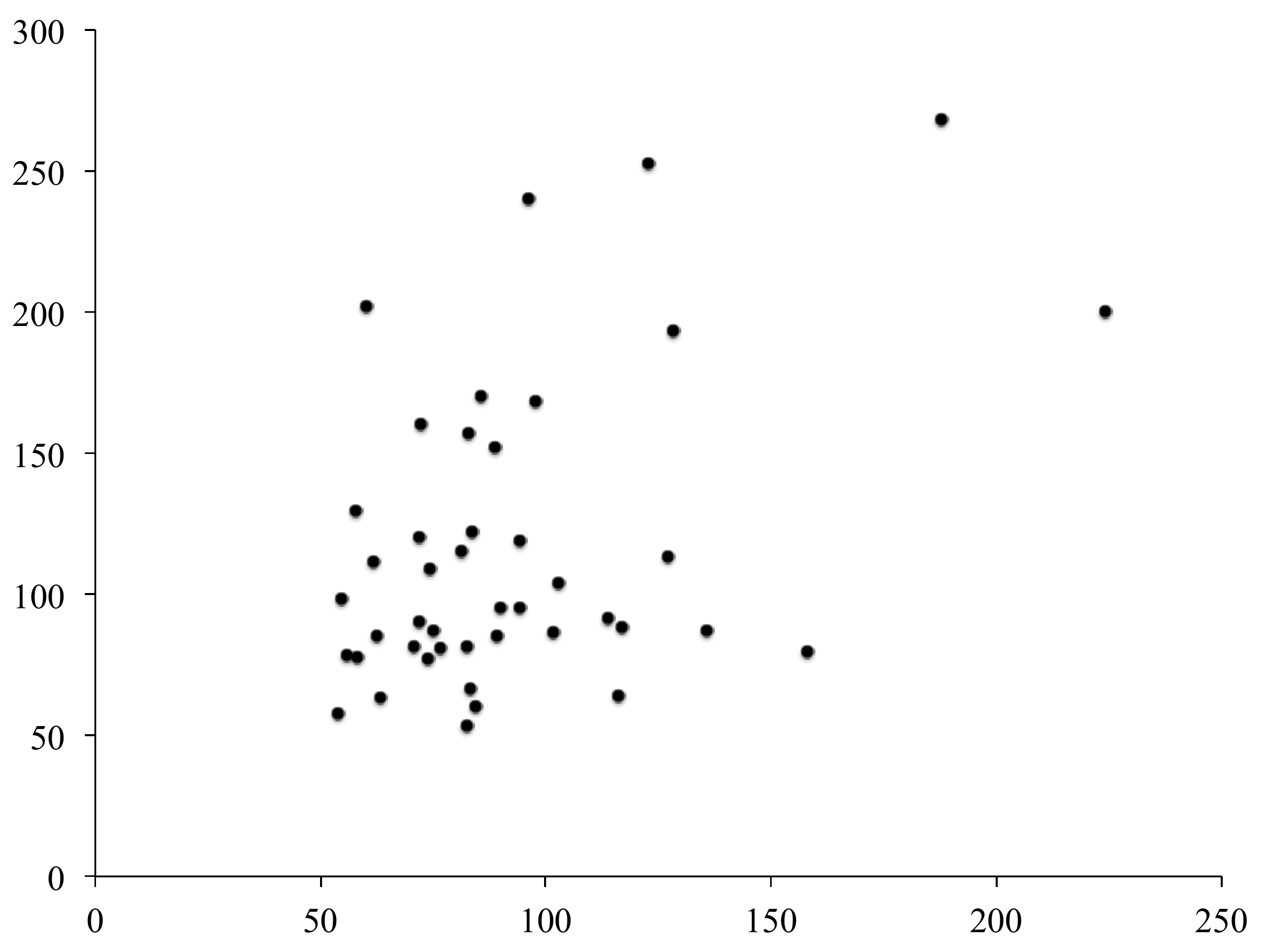}
    \caption{Scatter Plot from Stations $25081$ and $25078$}
    \label{Graph 3}
   	\end{center}
\end{figure}

The data points are arranged from oldest to newest, from left to right; therefore, the first value corresponds to the year of 1963 while the last value represents the year 2006.  In addition, tests were conducted using GNU R in order to assure that the data was homogeneous. First, the data was mapped using a scatter plot, Figure \ref{Graph 3} and a correlation test was performed, the p-value obtained was of $0.0031$, indicating that the data is significantly correlated. However, as seen in the theoretical framework in Section \ref{sec:theo}, practitioners should not be worried about putting together correlated data to perform the RAQE. Then, a paired-t test and a Levene test were conducted and the results indicated that they have different means and variances. In addition, confidence intervals using a boostrap with $1,000$ replicas were estimated for the skewness and the kurtosis. The results show that there is sufficient statistical evidence to state that their skewness and kurtosis are not significantly different.     

\begin{figure}[h]
	\begin{center}
    \includegraphics[width=.85\textwidth]{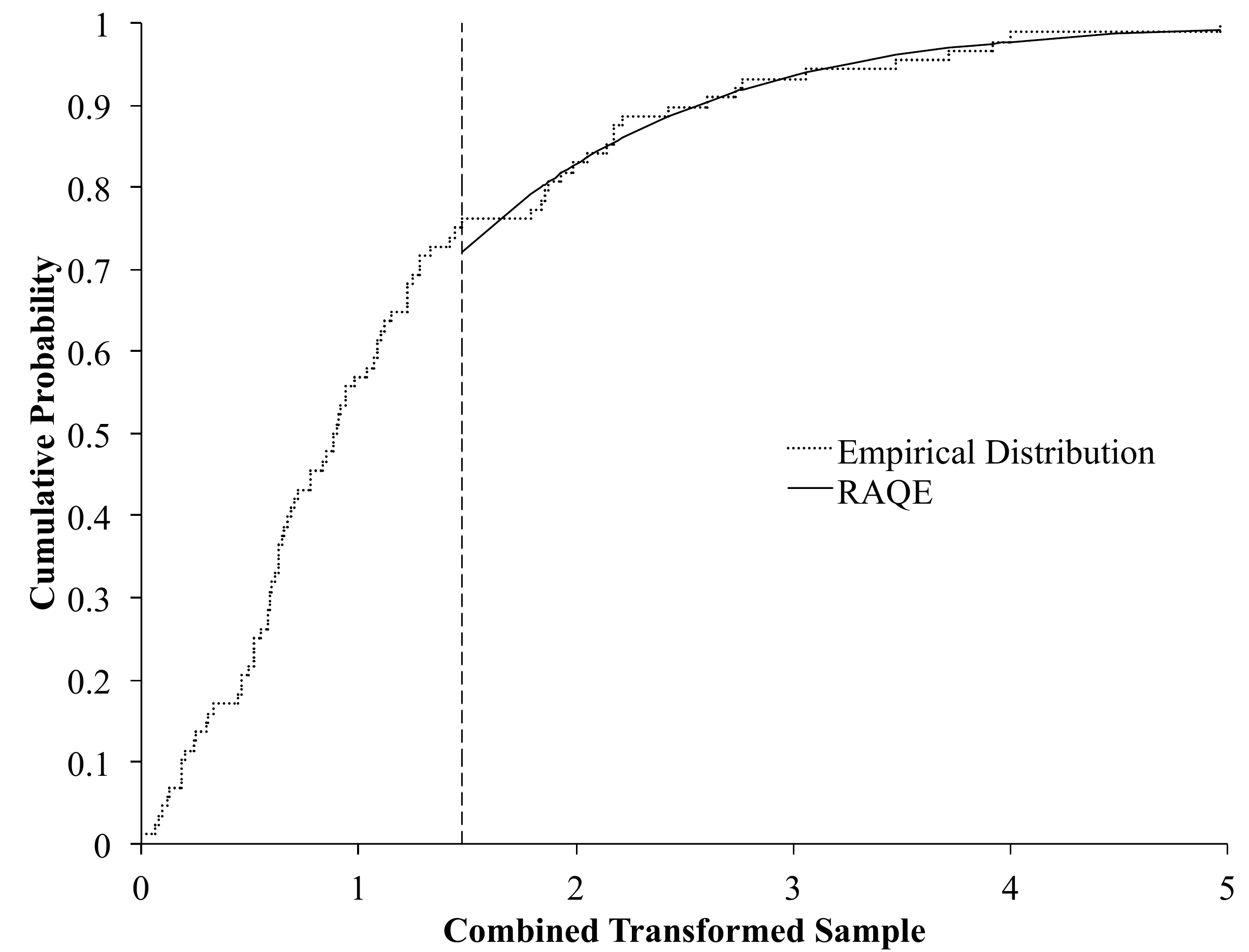}
    \caption{Sub-sample For the Upper Tail}
    \label{Graph 4}
    \end{center}
\end{figure}

After all the assumptions were tested, RAQE was estimated with m and l set on $25\%$ of the data. The selected model to estimate the return period was a Gumbel function, 
 $exp\left[-exp\left(-\left(x_i-\theta_1\right)/\theta_2\right) \right]$, this can be seen graphically in Figure \ref{Graph 4}. The return periods estimated were of $1,000$, $100$, and $20$ years, which correspond to the quantiles $0.999$, $0.99$, and $0.95$. The estimated quantiles for the station $25081$ were $295.031, 218.54, 164.51$; and for the station $25078$ the quantiles were $429.51, 311.14, 227.51$.

\section{Theoretical Framework}
\label{sec:theo}
The quantile estimation based on local curve fitting uses the theory of the empirical distribution. A modification of the empirical distribution function, the augmented empirical distribution, was proposed to be used with existing regression techniques to find the quantiles of interest by using its inverse image. These results were already presented by \cite{salazar2016regressing} in their previous work. Nevertheless, for completeness, each of these subjects is described next in this section together with the new theoretical developments. In addition, a discussion about correlation between samples is also presented. 

\subsection{The Empirical Distribution}

\textbf{Definition 1.}\emph{ Let $X_1,\ldots,X_n$ be a series of independent and identically distributed copies of a random variable $X$ with continuous distribution function $F(x)= P(X \leq x)$. The statistic
\begin{equation}
	S(x)=\frac{1}{n} \sum_{i=1}^{n} I_{(-\infty,x]}  (X_i)
\end{equation}
is defined as the empirical distribution function (e.d.f), for all $x$, $-\infty<x<\infty$.}

$I_{(-\infty,x]}(\cdot)$ is an indicator function that takes values of $1$ when evaluated in $(-\infty,x]$, and zero otherwise. An important properties of the empirical distribution is that $I_{(-\infty,x]}(X_i)$  is a Bernoulli random variable, hence $S(x)$ is the MLE of the cumulative probability $F(x)$. In addition, at any given value of $x$
\begin{equation}
E[S(x)]=F(x)                                                         
\end{equation} 
\begin{equation}                                 Var[S(x)]=\frac{1}{n} (F(x)(1-F(x)).                                       
\end{equation} 
 Also, the Gilvenko-Cantelli theorem,
 \begin{equation}
sup|S(x)-F(x)|\overset{a.s.}{\rightarrow} 0,                                                       
\end{equation} 
 states that $S(x)$ converges uniform to $F(x)$ as $n  \rightarrow \infty$.

As demonstrated by \cite{salazar2016regressing} the augmented empirical distribution estimates points from the true continuous cumulative probability distribution function. Hence, this recollection of points can be used to fit a curve, with a least squares procedure, to find a quantile of interest. The obtained estimates are unbiased.  However, \cite{salazar2016regressing} did not considered that different $b_i$, $ i=1, \dots, n$, are heteroscedastic where the variance can be estimated as $\hat{\sigma}_1^2=b_i(1-b_i)/n$. This can be addressed by adding weights of  $w_i=1/\hat{\sigma}_i^2$ to the least squares criterion, 

\begin{equation}
Q_w=\sum_{i=1}^{n} {w_i (b_i - \hat{b}_i)^2}.
\end{equation}

In addition, for the case of correlated data when using multiple homogeneous samples, suppose we have two sets of independent and identically distributed random variables  $X_1, \ldots, X_{n_1}$  and $Y_1, \ldots, Y_{n_2}$. Suppose we analyze each sample separately and the probability at a threshold $a$ is of interest. Then, for sample $1$
\begin{equation}
\hat{b}_{1i}= \frac{1}{n_1} \sum_{i=1}^{n_1}I(X_i \leq a)
\end{equation}     
Similarly, for sample $2$
\begin{equation}
\hat{b}_{2i}= \frac{1}{n_2}\sum_{i=1}^{n_2}I(Y_i \leq a)
\end{equation}  
In both cases, an unbiased estimator of $F(x)$ is obtained. If the two samples are now combined in order to obtained more information about the process,then,
\begin{align}
\begin{split}
\hat{b}_i=&\frac{n_1\hat{b_{1i}}+n_2\hat{b_{2i}}}{n_1+n_2}
\end{split}
\end{align} 
Since the probability is a linear combination of the individual estimators, then, it still is an unbiased estimator with variance equal to $\frac{1}{n_1+n_2}\theta (1-\theta) + \frac{2n_1n_2}{\left(n_1+n_2 \right)^2} Cov(\hat{b}_{1i}, \hat{b}_{2i})$, where $\theta=P(X \leq a)$.  Therefore, practitioners should feel comfortable even though the samples are not independent.

\section{Conclusions}
\label{sec:conc}
In the literature, several techniques exist to estimate quantiles by using an overall fit of the data available. However, to estimate extreme quantiles the best method would be the one that provides the best estimation at the tails of the distributions. RAQE is a new method that was presented to solve the problem of process capability analysis when the data does not represent a normal distribution. This method is now generalized so it can be used as a technique for quantile estimation.

RAQE was presented for its use with one sample or  multiple homogeneous samples. As well, real case examples were presented for the problem of estimation of control limits in quality control and the estimation of return periods in hydrology. In both cases, it is noticeable that the results are similar to the results obtained with other techniques. This technique provides practitioners with a new alternative to estimate extreme quantiles.  In some cases, the estimation even requires fewer steps to obtain the quantile of interest. 

As part of the future work, further analysis of this method can be done using Monte Carlo simulations; nevertheless, special considerations need to be made when capturing the behavior of the tails of different samples by using only a model. In practice, practitioners have a particular sample of interest and a particular curve can be proposed by examining the data set, and they will choose the approach that best fits their distribution using popular selection approaches such as the mean absolute deviation, BIC or AIC. This situation is similar to the one presented in this work.

\subsection{Acknowledgements}

\subsection{Funding}

\bibliographystyle{gNST}
\bibliography{references}

\begin{thebibliography}{13}
\newcommand{\enquote}[1]{`#1'}
\providecommand{\natexlab}[1]{#1}
\providecommand{\url}[1]{\normalfont{#1}}
\providecommand{\urlprefix}{ }

\bibitem[Arellano-Lara and Escalante-Sandoval(2014)]{lara2014multivariate}
Arellano-Lara, F.d.R., and Escalante-Sandoval, C.A. (2014),
  \enquote{Multivariate delineation of rainfall homogeneous regions for
  estimating quantiles of maximum daily rainfall: A case study of northwestern
  Mexico}, \emph{Atm{\'o}sfera}, 27.

\bibitem[Bartlett(1947)]{bartlett1947use}
Bartlett, M.S. (1947), \enquote{The use of transformations}, \emph{Biometrics},
  3, 39--52.

\bibitem[Box and Cox(1964)]{box1964analysis}
Box, G.E., and Cox, D.R. (1964), \enquote{An analysis of transformations},
  \emph{Journal of the Royal Statistical Society. Series B (Methodological)},
  pp. 211--252.

\bibitem[Chou et~al.(1998)Chou, Halverson, and Mandraccia]{chou1998control}
Chou, Y.M., Halverson, G.D., and Mandraccia, S.T. (1998), \enquote{Control
  charts for quality characteristics under nonnormal distributions},
  \emph{Statistical Case Studies: A Collaboration Between Academe and
  Industry}, 3, 89--97.

\bibitem[Daniel and Wood(1999)]{daniel1999fitting}
Daniel, C., and Wood, F.S. (1999), \emph{Fitting equations to data: computer
  analysis of multifactor data}, John Wiley \& Sons, Inc.

\bibitem[Davison and Hinkley(1997)]{davison1997bootstrap}
Davison, A.C., and Hinkley, D.V. (1997), \emph{Bootstrap methods and their
  application}, Vol.~1, Cambridge University press.

\bibitem[Efron(1982)]{efron1982jackknife}
Efron, B. (1982), \emph{The jackknife, the bootstrap and other resampling
  plans}, Vol.~38, SIAM.

\bibitem[Efron and Gong(1983)]{efron1983leisurely}
Efron, B., and Gong, G. (1983), \enquote{A leisurely look at the bootstrap, the
  jackknife, and cross-validation}, \emph{The American Statistician}, 37,
  36--48.

\bibitem[Field and Smith(1994)]{field1994robust}
Field, C., and Smith, B. (1994), \enquote{Robust estimation: a weighted maximum
  likelihood approach}, \emph{International Statistical Review/Revue
  Internationale de Statistique}, pp. 405--424.

\bibitem[Johnson(1949)]{johnson1949systems}
Johnson, N.L. (1949), \enquote{Systems of frequency curves generated by methods
  of translation}, \emph{Biometrika}, 36, 149--176.

\bibitem[Salazar-Alvarez et~al.(2016)]{salazar2016regressing}
Salazar-Alvarez, M.I., Temblador-P{\'e}rez, C., Conover, W.J.,
  Tercero-G{\'o}mez, V.G., Cordero-Franco, A.E., and Beruvides, M.G. (2016),
  \enquote{Regressing sample quantiles to perform nonparametric capability
  analysis}, \emph{The International Journal of Advanced Manufacturing
  Technology}, pp. 1--10.

\bibitem[Tan et~al.(2004)Tan, Gan, and Chang]{tan2004using}
Tan, W., Gan, F., and Chang, T. (2004), \enquote{Using normal quantile plot to
  select an appropriate transformation to achieve normality},
  \emph{Computational statistics \& data analysis}, 45, 609--619.

\bibitem[Velleman and Hoaglin(1981)]{velleman1981applications}
Velleman, P.F., and Hoaglin, D.C. (1981), \emph{Applications, basics, and
  computing of exploratory data analysis}, Duxbury Press.

\end{thebibliography}

\end{document}